\documentclass[12pt,preprint]{aastex}

\usepackage{epsfig}
\usepackage{natbib}                % To format bibliographies.

	\shorttitle{HD~37366, HD~54662}
	\shortauthors{Boyajian et al.}
\begin{document}

%%\received{}
%%\accepted{}

\title{The Long Period, Massive Binaries HD 37366 and HD 54662: \\
 Potential Targets for Long Baseline Optical 
Interferometry\altaffilmark{1}} 

\altaffiltext{1}{Based on observations obtained 
at the Canada-France-Hawaii Telescope (CFHT) which 
is operated by the National Research Council of Canada, 
the Institut National des Sciences de l'Univers of the 
Centre National de la Recherche Scientifique of France, 
and the University of Hawaii.}

\author{T. S. Boyajian\altaffilmark{2},  D. R. Gies\altaffilmark{2}, \\  
J. P. Dunn, C. D. Farrington, 
E. D. Grundstrom\altaffilmark{2},  W. Huang\altaffilmark{3}, \\ 
M. V. McSwain\altaffilmark{2,4,5}, S. J. Williams\altaffilmark{2}, D. W. Wingert\altaffilmark{2}}
\affil{Center for High Angular Resolution Astronomy and \\
 Department of Physics and Astronomy,\\
 Georgia State University, P. O. Box 4106, Atlanta, GA 30302-4106; \\
 tabetha@chara.gsu.edu, gies@chara.gsu.edu,
 dunn@chara.gsu.edu, farrington@chara.gsu.edu,
 erika@chara.gsu.edu, wenjin@astro.caltech.edu,
 mcswain@astro.yale.edu, swilliams@chara.gsu.edu, wingert@chara.gsu.edu}

\altaffiltext{2}{Visiting Astronomer, Kitt Peak National Observatory,
National Optical Astronomy Observatory, operated by the Association
of Universities for Research in Astronomy, Inc., under contract with
the National Science Foundation.}
\altaffiltext{3}{Current Address: Department of Astronomy, 
California Institute of Technology, MS 105-24, Pasadena, CA 91125} 
\altaffiltext{4}{Current Address: Astronomy Department,
Yale University, New Haven, CT 06520-8101}
\altaffiltext{5}{NSF Astronomy and Astrophysics Postdoctoral Fellow}

\author{A. W. Fullerton}     
\affil{Space Telescope Science Institute, 
3700 San Martin Drive, Baltimore, MD 21218; \\
fullerton@stsci.edu}

\author{C. T. Bolton}
\affil{David Dunlap Observatory, University of Toronto\\
P.O.\ Box 360, Richmond Hill, Ontario, L4C 4Y6, Canada;\\
bolton@astro.utoronto.ca}

%%\slugcomment{}
%%\paperid{}
%%%%%%%%%%%%%%%%%%%%%%%%%%%%%%%%%%%%%%%%%%%%%%%%%%%%%%%%%%%%%%
%DEFINITIONS FOLLOW
%water
\def\hzo        {H$_2$O}
%km/s
\def\kms    {\ifmmode{{\rm km~s}^{-1}}\else{km~s$^{-1}$}\fi}
%Mdot
\def\Mdot   {\ifmmode {\dot M} \else $\dot M$\fi}
%solar masses per year
\def\Mspy   {\ifmmode {M_{\odot} {\rm yr}^{-1}} \else $M_{\odot}$~yr$^{-1}$\fi}
%solar mass
\def\Msun   {$M_{\odot}$}
%micron
\def\mum     {\ifmmode{\mu{\rm m}}\else{$\mu{\rm m}$}\fi}
%R_star
\def\Rstar  {$R_{\star}$}
%
%definitions end
%%%%%%%%%%%%%%%%%%%%%%%%%%%%%%%%%%%%%%%%%

\begin{abstract}

We present the results from an optical spectroscopic analysis of 
the massive stars HD~37366 and HD~54662.  We find that 
HD~37366 is a double-lined spectroscopic binary with a period of 
$31.8187\pm0.0004$ days, and HD~54662 is also
a double lined binary with a much 
longer period of $557.8\pm0.3$ days. 
The primary of HD~37366 is classified as O9.5~V, and it 
contributes approximately two-thirds of the optical flux. 
The less luminous secondary is a broad-lined, 
early B-type main-sequence star.  
Tomographic reconstruction of the individual 
spectra of HD~37366 reveals
absorption lines present in each component, 
enabling us to constrain the nature of the secondary 
and physical characteristics of both stars.  Tomographic
reconstruction was not possible for HD~54662; however, 
we do present mean spectra from our observations that   
show that the secondary component is approximately half 
as bright as the primary.  The observed spectral energy 
distributions (SEDs) were fit with model SEDs and galactic 
reddening curves to determine the angular sizes of the stars.
By assuming radii appropriate for their classifications, 
we determine distance ranges of 1.4 -- 1.9 and 1.2 -- 1.5 kpc 
for HD~37366 and HD~54662, respectively.
\end{abstract}

%%%%%%%%%%%%%%%%%%%%%%%%%%%%%%%%%%%%%%%%%%%%%%%%%%%%%%%%%%%%%%%

\setcounter{footnote}{5}

\section{Introduction}                              % Section 1

There remains considerable uncertainty about 
the masses of the most massive stars because of the 
relatively small number of known binary systems for 
which accurate masses can be determined \citep{gie03}.
Spectroscopic measurements alone yield mass functions 
dependent on the unknown orbital inclination, 
and the determination of inclination requires either the 
good fortune of finding eclipsing binaries or the angular 
resolution of the orbit on the sky.  The angular semimajor axis 
of a binary (in units of milliarcsec = mas) is given by
\begin{equation}
a({\rm mas}) = 0.28 \frac{(P/(10~{\rm d}))^{2/3} (M_{\rm total}/(30 M_\odot ))^{1/3}}{  (d/(1 {\rm kpc}))}
\end{equation}
where $P$ is the orbital period, $M_{\rm total}$ is the 
combined mass of the stars, and $d$ is the distance.  
The denominators of each unit give typical values for these 
parameters among OB binaries, and the leading coefficient of
0.28~mas indicates that most massive systems are probably 
too closely separated for direct resolution with optical long baseline
interferometers, where the limits are currently above 1~mas.  
The key objective here is to find double-lined 
spectroscopic binaries with long orbital periods.  Such binaries
are difficult to detect because their orbital semiamplitudes 
are small, the component lines are often blended, and a 
long term observational program is required to obtain adequate 
phase coverage.  The best candidates for direct resolution are 
15~Mon ($P\approx 25$~y; \citealt{gie97}), 
HD~15558 ($P=442$~d; \citealt{gar81, deb06}), and
HD~193322 ($P=311$~d; \citealt{mck98}).   

Here we report on new orbits for two such long period 
massive binaries, HD~37366 and HD~54662.  
The star HD~37366 (BD$+30^\circ 968$; HIP 26611; 
O9.5~V, \citealt{wal73}) is a member of the Aur~OB1 association 
at a distance of approximately 1.3~kpc \citep{hum78}.
This association has many bright early-type giants 
and supergiants, but HD~37366 has the earliest spectral 
type among the member stars that still reside on the main sequence.
The {\it Hipparcos} mission \citep{per97} detected a visual 
companion to HD~37366 with $\Delta H_{p}=3.5$, a 
separation of $0\farcs58$, and a period of 
approximately 1300 years \citep{mas98}.  The brighter 
component of these two stars ($H_{p}=7.7$) 
is a radial velocity variable \citep{pet61,you42}, 
and it is known to show asymmetry in its spectral 
lines \citep*{gri92}.  Observations with the 
{\it International Ultraviolet Explorer (IUE)} confirm 
that the spectrum displays double lines \citep{sti01}. 

The second target HD~54662 (BD$-10^\circ 1892$; HIP 34536; 
LS 197; O6.5~V, \citealt{wal72}) is also the brightest 
and earliest member of its resident association, 
CMa~OB1, at a distance of 1.3 kpc \citep{hum78}.  
Radial velocity measurements for HD~54662 extend
back many decades \citep*{pla24,con77a,gar80}, 
and these display only modest variability. 
However, \citet{ful90} noted the presence of 
blue extensions to the spectral lines that 
probably indicate the presence of a companion 
in a long period orbit.  The scatter in the 
{\it IUE} velocities also indicates that the 
star is a binary \citep{sti01}. 

Here we present an analysis of the radial velocities 
and spectra of both stars from spectroscopic observations 
that we have obtained over the past few years (\S2). 
We discuss each system's orbital velocity solution (\S3)  
and the spectral and physical properties of each component 
star in these binaries (\S4,5).  We conclude with a 
consideration of the prospects for the angular resolution of the 
orbits using optical long baseline interferometry (\S6).

%%%%%%%%%%%%%%%%%%%%%%%%%%%%%%%%%%%%%%%%%%%%%%%%%%%%%%%%%%%%%%%

\section{Observations}        % Section 2

We observed HD~37366 and HD~54662 with the 
Kitt Peak National Observatory (KPNO) 
0.9~m Coud\'{e} Feed telescope during 
two separate observing runs in 2000 
October and 2000 December.  
The spectra were made using the 
long collimator, grating B (in second order with order sorting 
filter OG 550), camera 5, and the F3KB CCD, a Ford Aerospace 
$3072 \times 1024$ device  with $15 \mu$m square pixels.  
The setup yielded a resolving 
power of $R=\lambda/\delta\lambda$=9500, with a spectral 
coverage of 6440$-$7105 \AA.  Exposure times were 
usually 10 minutes or less, and we generally obtained
two spectra (taken a few hours apart) each 
night. For HD~37366, we made 
two more red spectral observations in 2004 October
using a similar arrangement but with a different 
detector, the T2KB CCD ($2048 \times 2048$ $24 \mu$m square pixels). 
In 2006 October, both HD~37366 and HD~54662 
were observed in the red region
again using this same instrumental set-up.
We also observed the rapidly rotating A-type star,
$\zeta$~Aql, which we used for removal of 
atmospheric water vapor and O$_2$ bands.
Each set of observations was accompanied by 
numerous bias, flat field,
and Th-Ar comparison lamp calibration frames.

We also obtained a small set of blue spectra of these targets. 
For HD~37366, the first group of four 
spectra were made in 2005 October with 
the KPNO 2.1~m telescope and GoldCam spectrometer.
We used the $\#47$ grating in second order, 
recording the spectral region from 4050$-$4950\AA ~with a 
resolving power of $R=\lambda/\delta\lambda\approx 3000$.
Then in 2005 November and 2006 October we obtained 
higher resolution observations in the blue with 
the KPNO Coud\'{e} Feed 0.9~m telescope. 
HD~37366 was observed on both occasions, 
whereas HD~54662 was only included during the 2006 observing run.  
We used grating A in second order
with order sorting filter 4-96, camera 5, and the T2KB CCD.
This setup gave us a resolving power of 
$R=\lambda/\delta\lambda\approx 12100$ and a wavelength 
coverage of 4240$-$4585\AA.  

The spectra were extracted and calibrated
using standard routines in IRAF\footnote{IRAF is distributed by the
National Optical Astronomy Observatory, which is operated by
the Association of Universities for Research in Astronomy, Inc.,
under cooperative agreement with the National Science Foundation.}.
All the spectra were rectified to a unit continuum by fitting
line-free regions.  The removal of atmospheric lines 
from the red spectra was done by
creating a library of $\zeta$~Aql spectra from each run, removing
the broad stellar features from these, and then dividing each target
spectrum by the modified atmospheric spectrum that most closely
matched the target spectrum in a selected region dominated by
atmospheric absorptions.  The spectra from each run were then
transformed to a common heliocentric wavelength grid.
   
%%%%%%%%%%%%%%%%%%%%%%%%%%%%%%%%%%%%%%%%%%%%%%%%%%%%%%%%%%%%%

\section{Radial Velocities and Orbital Elements}   %Section 3

\subsection{HD~37366}

We measured radial velocities of the 
high resolution red spectra 
collected in 2000 and 2006 using 
a template-fitting scheme \citep{gie02} 
for the \ion{He}{1}~$\lambda 6678$ line.
We decided not to measure the other strong lines  
in this region because the binary components 
are badly blended in the H$\alpha$ profile and the  
\ion{He}{1}~$\lambda 7065$ line was marred by 
residual features left behind by the telluric 
cleaning procedure.  
This radial velocity measurement scheme assigns template spectra that
are approximate matches for the primary
(hotter and more massive star) and secondary
spectra, and then makes a non-linear least-squares fit 
of the shifts for each component that best match 
the observed line profile.  We need to make assumptions 
at the outset about the temperature, gravity, projected 
rotational velocity, and flux contribution of each star, 
but these can be checked after completion of the 
velocity analysis by studying the properties of 
tomographically reconstructed spectra of the components (\S4). 

The matching template spectra for the primary and secondary 
components were constructed from the grid of O-type star model 
spectra from \citet{lan03} that are based upon the line blanketed, 
non-LTE, plane-parallel, hydrostatic atmosphere code {\it TLUSTY} 
and the radiative transfer code {\it SYNSPEC} \citep*{hub88,hub95,hub98}.  
We selected the spectrum taken on HJD~2,451,901.92, 
which shows well separated, individual components of 
each star, as a reference to determine 
the approximate spectral parameters for both stars. 

The template fitting procedure also requires
preliminary estimates of the primary and secondary 
stars' radial velocities.  We estimated these for each spectrum
with well separated lines using the IRAF {\it 
splot} routine and deblend option 
to fit two Gaussians to each composite profile. 
We also measured relative radial velocity shifts 
of the strong interstellar lines in all the spectra 
referenced to the first spectrum in the stack.  We then 
used these relative shifts in the interstellar lines 
(which should remain motionless) to make additional 
small corrections for the wavelength calibrations 
(all these corrections were $<2$ \kms).  

The final radial velocities from this template 
fitting procedure (the majority of the observations)
are listed in Table~1 along with the
heliocentric Julian date of mid-observation, 
the corresponding orbital phase,  
and the residual from the orbital fit
(observed minus calculated) for both the primary and 
the secondary.  The typical errors in these velocities
are also listed in Table~1.  We measure only one line for this data set, 
so we list the characteristic 
errors (not individual errors), which are based upon 
the scatter in closely spaced pairs 
of observations. These errors are 1.3 and 2.2 
km~s$^{-1}$ for the primary and secondary, respectively.
  
\placetable{tab1}      % Table 1 - Radial velocities for HD 37366

This template fitting routine was also used in 
determining radial velocities for the high resolution
blue spectra (collected in 2005 November 
and 2006 October), using 
the four lines \ion{O}{2}~$\lambda 4349$, 
\ion{He}{1}~$\lambda\lambda 4387, 4471$,
and \ion{Mg}{2}~$\lambda 4481$.  We followed 
the same procedure in obtaining the spectral templates
and the preliminary radial velocity
estimates as described above.  Since no 
strong interstellar features are apparent in this region, 
no additional radial velocity correction was applied.  
The line-to-line $1\sigma$ errors in 
these $V_{R}$ measurements are $< 1$ 
\kms~for the primary, and
$4-9$ \kms~for the secondary (Table~1).  These final $V_{R}$  
measurements are also presented in Table~1.  
    
The four observations made in the blue during  
2005 October had a much lower resolution, thus making it 
difficult to apply this method of template fitting.  
To avoid possible errors from unseen line blending
in the two components, we chose to measure only radial 
velocities of the \ion{He}{2}~$\lambda\lambda 4541, 4686$
lines present since these 
lines are found only in the spectrum of the 
much hotter, primary star (\S4).  
We used a parabolic fitting routine
to determine the mean velocities of these lines (Table~1).  The 
line-to-line $1\sigma$ errors associated with these 
measurements are $< 6$ \kms (Table~1). 

The two red observations made in 2004 showed no
indication of double-lined profiles.  In this case, 
we measured velocities only for the primary 
star by parabolic fitting of the line cores of \ion{He}{1}
$\lambda\lambda 6678, 7065$, in order to minimize the
influence of the secondary on the line profile.  
The line-to-line $1\sigma$ error associated 
with these fits are $< 2$ \kms~(exclusive of blending 
errors). These velocities are also presented in Table~1.

The final two spectra of HD~37366 were collected and 
downloaded from the archive of the 
International Ultraviolet Explorer ({\it IUE}) 
satellite\footnote{http://archive.stsci.edu/iue/}.  
We measured radial velocities for these two high dispersion, 
short wavelength, prime camera spectra using a 
cross-correlation method \citep{pen99} with 
the spectrum of HD~34078 
as the reference template. The spectrum was 
double-lined in the first spectrum, SWP~30165.  
The errors are approximately 5~\kms~ 
for the primary and 10~\kms~for the 
secondary.  These final 
velocities are also presented in Table~1.

The radial velocities from all 
the data sets (six total) 
span 20 years with 45 radial velocity 
measurements for the primary and
38 radial velocity measurements for the secondary (Table~1).
We first constructed a power spectrum using all 
the primary star's radial velocity
measurements, being more reliable and plentiful, 
to identify possible orbital periods for 
the binary.  We
used the discrete Fourier
transform and CLEAN deconvolution algorithm \citep*{rob87}, 
which shows that the strongest signal occurs near 
$P = 31.7$ days. We then used this estimate as 
a starting value for the period in  
fits of the orbital elements.

We determined the orbital elements 
of the binary using the non-linear, least-squares,
orbital fitting 
program from \citet{mor74}.  We began with a 
fit of the primary's velocities that is
given in column~2 of Table~2 done with equal
weighting except for the low resolution blue
spectra, which have a weight set to zero.  This solution 
has a period of $31.8187 \pm 0.0004$ days. 
The independent orbital solution for 
the secondary has a period of 
$31.822 \pm 0.002$ days, given in 
column~3 of Table~2.  Since the independent solutions 
agree well with each other, we derive a joint 
solution by fixing the weighted means of the shared 
orbital parameters ($P$, $T$, $e$, $\omega$) found in the 
independent solutions for the binary
in order to make fits of the systemic 
velocity, $\gamma_{1,2}$,
and the semiamplitude, $K_{1,2}$, for 
each component (column~4 of Table~2).
In the case of massive binaries, the systemic 
velocities of the components may not 
agree exactly because of differences in their expanding 
atmospheres and/or in our case, differences in the shapes of 
the template spectra for the \ion{He}{1}~$\lambda6678$,
\ion{He}{2}~$\lambda6683$ blend.  The radial velocity curves for the 
joint solution are plotted together 
with the observations in Figure~1.  We also made 
similar fits weighting each each point by the normalized,
inverse square of its associated error, these results 
matched within errors of those from the equal weighting fits 
given in Table~2.

\placetable{tab2}      % Table 2 - Orbital elements for HD 37366

\begin{figure}
\begin{center}
{\includegraphics[angle=90,height=12cm]{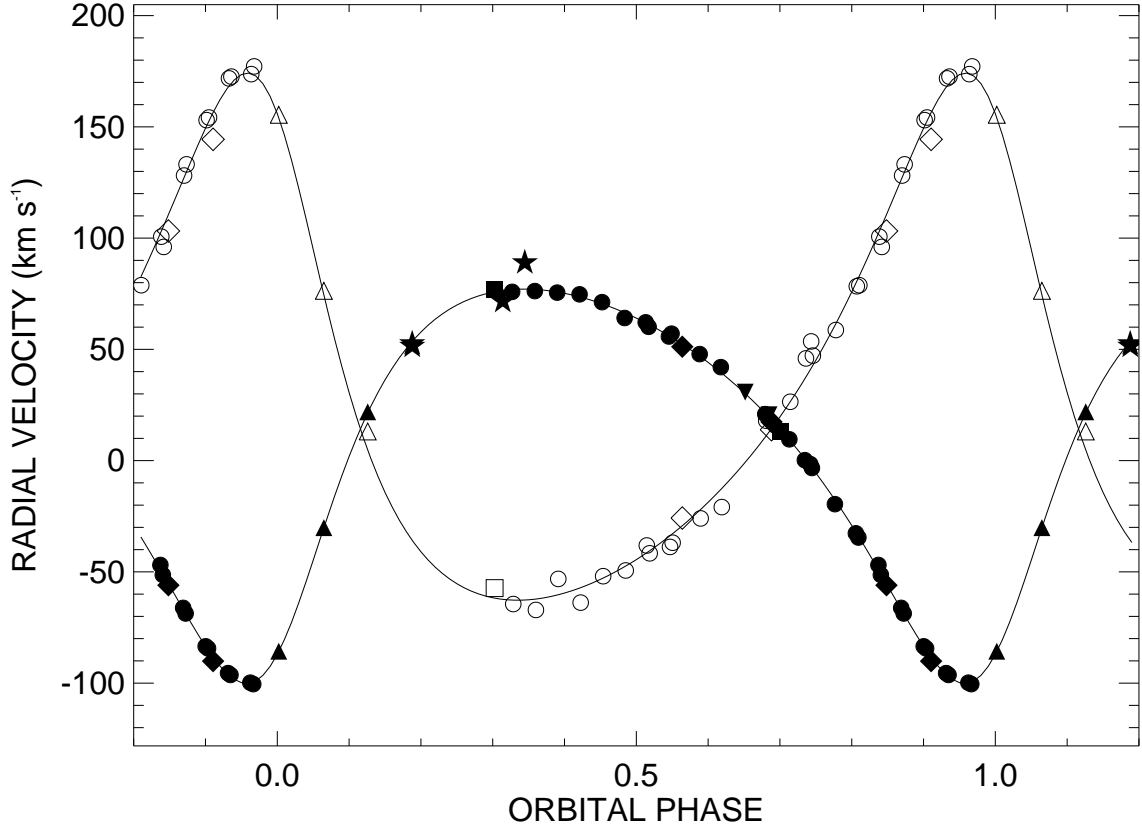}}
\end{center}
\caption{Calculated radial velocity curves 
({\it solid lines}) for HD~37366.  
The primary and secondary star's measured 
radial velocities are indicated by {\it circles} 
(2000), {\it inverted triangles} (2004 October), 
{\it stars} (2005 October), {\it triangles} 
(2005 November), {\it diamonds} 
(2006 October), and {\it squares} ({\it IUE} 
1987).  The filled symbols correspond to the 
primary and the open symbols to the secondary.  
The uncertainties in individual measurements
are generally smaller than the size of the symbols.}
\label{fig1}
\end{figure}

\placefigure{fig1}     % Figure 1 - Radial velocity curve for HD 37366

\subsection{HD~54662}		% Section 3.2

We obtained relative radial velocities for the 
primary (hotter, more massive) star in HD~54662 by
cross-correlation with a single spectrum
of the star that had good S/N properties. 
These relative velocities were 
transformed to an absolute velocity 
scale by adding the mean
velocity measured through parabolic 
fits to the cores of the absorption 
lines in this reference spectrum.  
All the strong lines were included
in the cross-correlation measurements, 
namely, H$\alpha$, \ion{He}{1}~
$\lambda6678$ $+$ \ion{He}{2}~
$\lambda6683$, and \ion{He}{1}~
$\lambda7065$.  We excluded   
\ion{He}{2}~$\lambda\lambda 6527, 6890$  
because their measurements deviated 
from the set listed above, as well 
as compared to each other. We suspect that the residual
telluric lines in the spectra, that are very
prominent in these regions, are the cause of this
disagreement. The velocities for the primary star
are presented in Table~3, along with the 
average velocity and $\sigma$ (line-to-line) 
from the \ion{C}{4}~$\lambda\lambda5801,5812$ 
and \ion{He}{1}~$\lambda5876$
lines presented by \citet{ful90}. 
We were unable to measure velocities for the 
secondary star in individual spectra due to severe 
line blending with profiles of the 
primary star (see \S5). 

Published velocities for HD~54662 
(\S1) do not show significant 
variations. Table~3 shows that 
our measurements change only slightly 
over our observation period.  
However, \citet{ful90} 
found convincing evidence that 
this system is a double-lined binary with 
either a long period or high eccentricity, 
since he observed a blue-shifted secondary component 
(suspected ~O7 spectral type) in the profiles of 
\ion{C}{4}~$\lambda\lambda 5801, 5812$
and \ion{He}{1}~$\lambda 5876$.

Here we present a preliminary orbital
solution for the primary component that was determined 
using our measurements combined with published
measurements \citep{pla24,gar80,ful90,sti01} for a total
of 67 radial velocities spanning 85 years.   
\citet{sti01} proposed a possible 
period of $\approx 92$~d, however, their orbit
was determined excluding selected data points.
We re-investigated the possible period by 
power spectrum analysis of all the available data.  
We examined all the peaks in the CLEANed 
spectrum using the non-linear, least-squares, 
orbital fitting routine, and among the 
periods limited by the timescales sampled in
our two long runs, we find 
that the best solution occurs at a period 
of $\approx$558 days. 
This confirms the suggestion from 
\citet{ful90} that HD~54662 is in fact a
long period binary.  
Table~4 lists the preliminary orbital
elements for HD~54662 assuming equal 
weighting for all velocities, 
and this solution is plotted in Figure~2.  We 
show below (\S5) that these results are 
affected by line blending, and the derived semiamplitude, 
for example, is a lower limit to the actual value.
It is also possible that the results collected in the 
literature have systematic differences related to 
the specific lines and measurement techniques used.
These systematic offsets are likely much smaller than the 
system semi-amplitude, and since this system has such a 
long orbital period, we include all available 
measurements for this preliminary orbital solution.

\placetable{tab4}      % Table 4 - Orbital elements for HD54662

\begin{figure}
\begin{center}
{\includegraphics[angle=90,height=12cm]{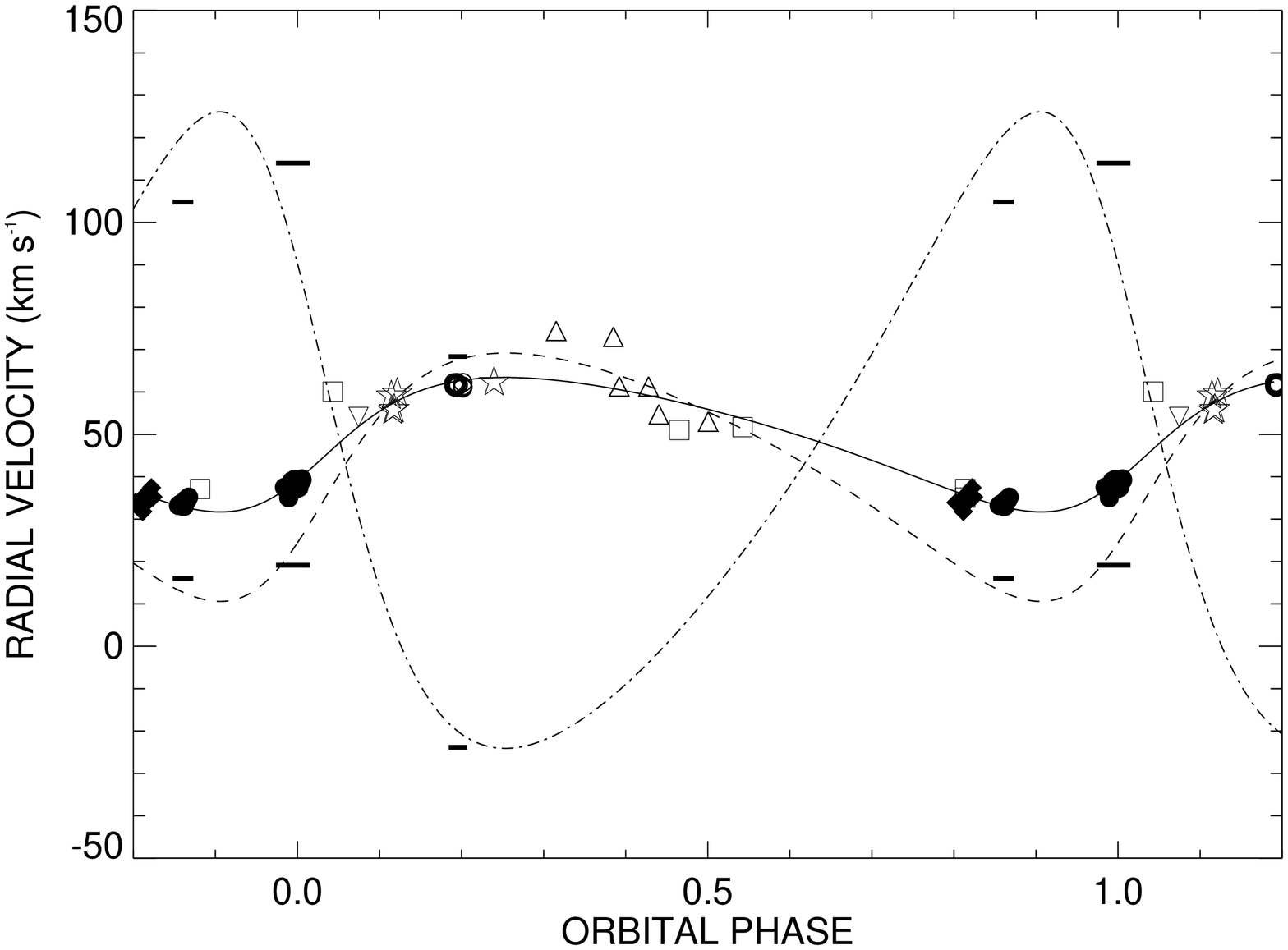}}
\end{center}
\caption{Tentative radial velocity curve ({\it solid 
line}) for HD~54662 for a period of 558~d.  
The measured radial velocities are indicated by 
{\it solid circles} (2000), {\it solid diamonds} 
(2006), {\it squares} (\citealt{sti01}), 
{\it open circles} (\citealt{ful90}), 
{\it stars} (\citealt{gar80}), {\it inverted triangles}
(\citealt{con77a}), and {\it triangles} (\citealt{pla24}).
Expanded {\it horizontal bars} are plotted to show the radial 
velocities derived from fitting the composite line profile 
from the average spectra for three observational epochs (\S5).
The {\it dashed} and {\it dot-dashed} lines 
are the radial velocity curves
fit to these time-averaged points for 
the primary and secondary star, respectively.   
The uncertainties in individual measurements
are generally smaller than the size of the symbols.  }
\label{fig2}
\end{figure}

\placefigure{fig2}     % Figure 2 - Radial velocity curve for HD54662

%%%%%%%%%%%%%%%%%%%%%%%%%%%%%%%%%%%%%%%%%%%%%%%%%%%%%%%%%%%%%%%

\section{Tomographic Spectral Reconstruction and Stellar Parameters for HD~37366}       % Section 4

We used a Doppler tomography algorithm
\citep{bag94} to separate the primary and
secondary spectra of HD~37366.  We applied
tomographic reconstruction to the red spectra
collected in 2000 (30 total) and to the high 
dispersion blue spectra collected in 2005 
and 2006 (five total).  Figure~3 shows the 
reconstructed red spectra for the 
primary ({\it top}) and the 
secondary ({\it bottom}).  The region affected by 
the atmospheric band from $\approx 6850-7000$ 
\AA~ was set to unity.  The secondary 
spectrum shows the weak lines of    
\ion{O}{2} $\lambda\lambda 6641, 6721$, and  
\ion{C}{2} $\lambda\lambda 6578, 6582$. These
lines are absent in the primary 
spectrum that shows instead features
such as \ion{He}{2} $\lambda 6683$ that 
are found in O-type spectra.  To determine 
a monochromatic flux ratio, $F_{2}/F_{1}$, we used the 
equivalent width of \ion{He}{1} $\lambda 6678$, 
since it does not change significantly 
with spectral type for late O- to 
early B- stars \citep{con74}. 
These equivalent widths 
in the primary and secondary 
reconstructed spectra are equal 
for a flux ratio of $F_{2}/F_{1} = 0.35 \pm 0.05$.       

\begin{figure}
\begin{center}
{\includegraphics[angle=90,height=12cm]{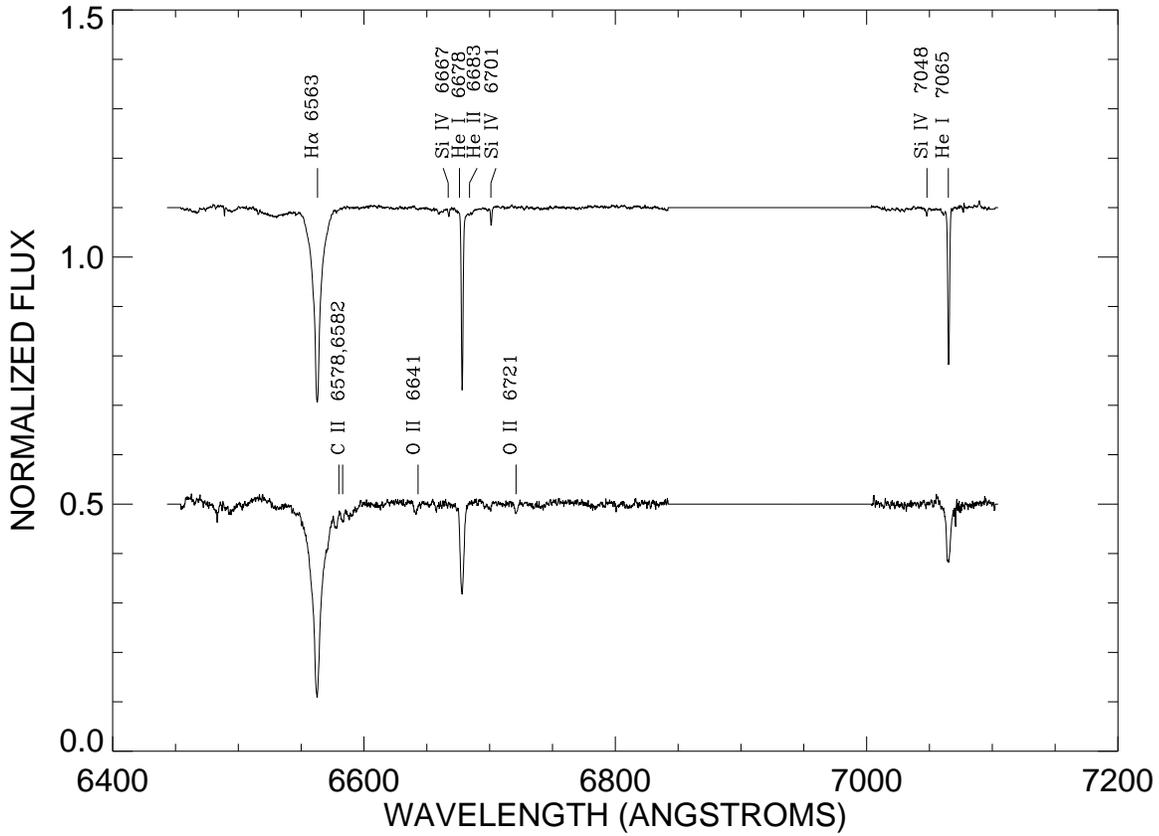}}
\end{center}
\caption{Tomographic reconstruction of the spectra of 
HD~37366 based upon the 30 red 
spectra obtained in 2000.  This plot shows the primary ({\it 
top}) and the secondary spectrum ({\it bottom}), as well as 
absorption line identifications 
({\it vertical marks}).  The atmospheric lines in the region of 
6850$-$7000\AA ~ are replaced with the continuum.}
\label{fig3}
\end{figure}

\placefigure{fig3}     % Figure 3 - TR mean spec of HD37366 in red

We fit these reconstructed spectra with 
the TLUSTY/SYNSPEC model synthetic spectra (see \S3) to 
estimate the projected rotational velocity
$V \sin i$, effective temperature 
$T_{\rm eff}$, and gravity $\log g$. 
These values are listed for both 
components of HD~37366 in Table~5 (where 
subscript 1 identifies the primary and 2 
the secondary).  For stars like these, the 
disappearance of the \ion{C}{2} and \ion{O}{2} 
lines and the emergence of the \ion{He}{2}
and \ion{Si}{4} lines with increasing 
temperature provide a useful temperature
estimate, while the width of the H$\alpha$ wings
is sensitive to the adopted gravity.  
$V \sin i$ was measured using 
a rotational broadening function applied to 
the model spectra to fit the two 
\ion{He}{1} absorption lines.  The red 
spectra were first used in the determination of 
these parameters, and the results were 
later checked with the reconstructed blue 
spectra, which include H$\gamma$ as well other 
lines from heavier elements. 
The small $V \sin i$ estimate we derive for the primary 
agrees with the {\it IUE} measurements from 
\citet{how97} and \citet{sti01} and is much smaller
than the value for the broader-lined secondary.
The primary's temperature is somewhat larger than
the $T_{\rm eff}=29.0 \pm 1.8$~kK 
estimate by \citet{gri92} but the gravities
agree exactly. Our results for $T_{\rm eff}$ 
and $\log g$ using the TLUSTY code
are expected to be more reliable than the  
previous models used in \citet{gri92}, which 
used the PAM code \citep{and85} that only includes 
nine elements and many fewer metal
lines than does TLUSTY.

The reconstructions from the five
high-resolution blue spectra 
are presented in Figure~4 along with  
identifications of absorption lines.
The secondary spectrum ({\it bottom}) has lower S/N, 
but even with only five spectra the tomography 
algorithm was able extract its spectrum. 
It is again apparent that the lines of the secondary 
are much broader than those of the 
primary.  Notice also the absence 
of the \ion{He}{2}~
$\lambda 4541$ line in the secondary's spectrum, 
reinforcing our conclusion that the secondary 
is the cooler of the two stars.  Based upon  
the secondary's cooler temperature and high surface 
gravity, we estimate that it is a B0$-$1~V star.  
Note that the magnitude difference we derive is 
larger than expected for main 
sequence stars separated by only a 
subtype or so \citep*{mar05}, so it is possible
that the primary is a somewhat evolved, 
more luminous star and/or the companion is a 
very young star close to the ZAMS.
It is interesting to note that 
the high $M_1 \sin^3 i$ and
$M_2 \sin^3 i$ values from the orbital solution 
suggest that the inclination
is large, $i = 60-90\degr$.  However, {\it Hipparcos} 
photometry plotted with the period from our
spectroscopic orbital solution shows no 
evidence of eclipses.

\begin{figure}
\begin{center}
{\includegraphics[angle=90,height=12cm]{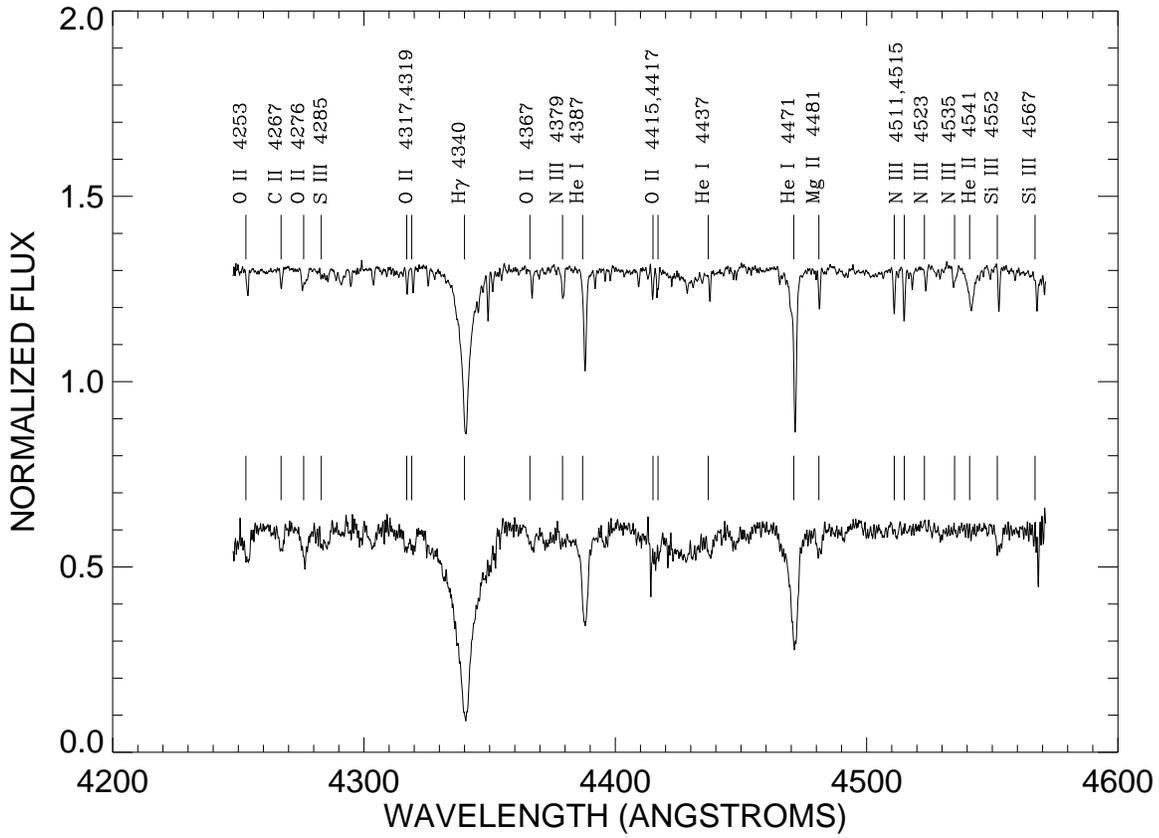}}
\end{center}
\caption{Tomographic reconstruction of the spectra of
HD~37366 based upon five  
blue spectra from runs in 2005 and 2006.
This plot shows the primary ({\it top}) and the secondary spectrum  
({\it bottom}) as well as a number of absorption 
line identifications ({\it vertical marks}).}
\label{fig4}
\end{figure}

\placefigure{fig4}	% Figure 4 - TR mean spec of HD37366 in the blue

\placetable{tab5}	% Table 5 - Stellar parameters of HD 37366

%%%%%%%%%%%%%%%%%%%%%%%%%%%%%%%%%%%%%%%%%%%%%%%%%%%%%%%%

\section{Stellar Parameters for HD~54662 from Composite Profile Fits}  % Section 5

Radial velocities measured for HD~54662 were 
used to create mean spectra for our observations
made in 2000 (Fig.~5) and for those 
made by \citet{ful90} in 1986.
Figure~6 shows an expanded view 
of the regions surrounding the \ion{He}{1}
profiles for two epochs of observation.  We see that 
the secondary component appeared blue-shifted 
during the 1986 run ({\it left panel}) and 
red-shifted in recent spectra ({\it right panel}).  

\begin{figure}
\begin{center}
{\includegraphics[angle=90,height=12cm]{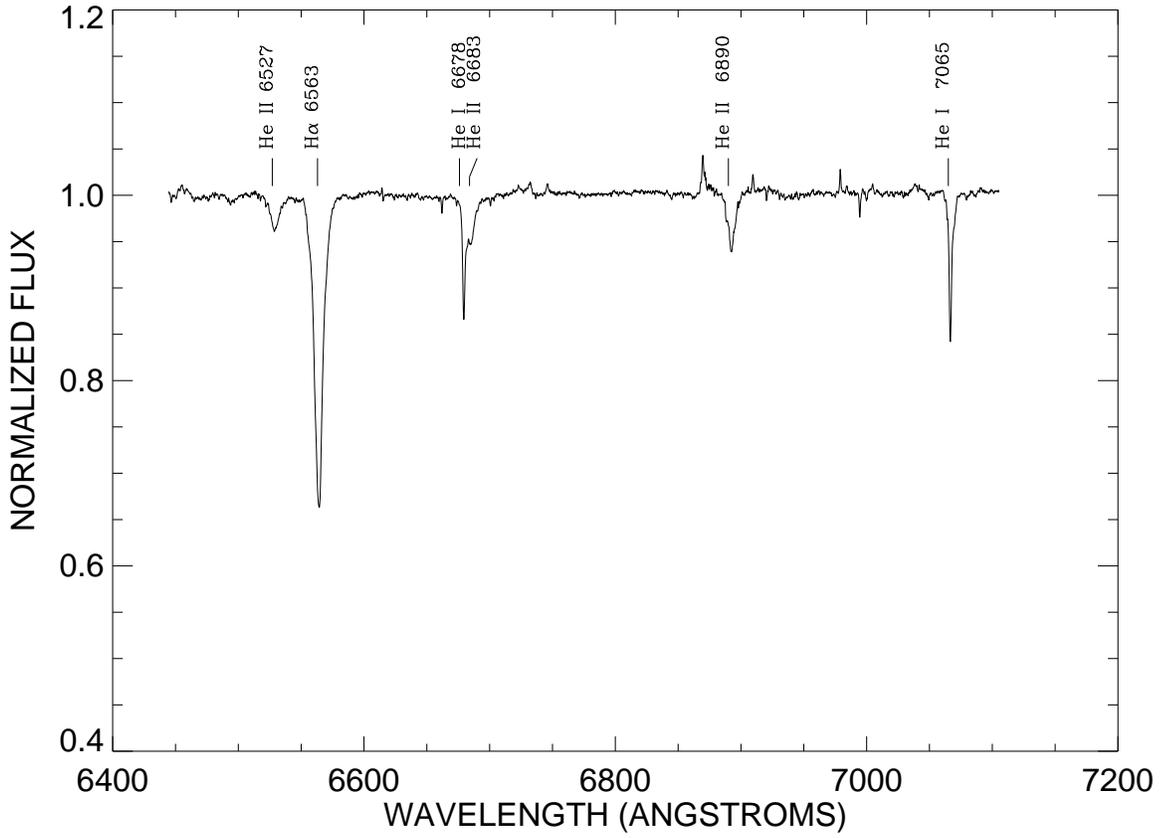}}
\end{center}
\caption{Mean spectrum of HD~54662 from our observations.
Line identifications are marked by vertical lines.}
\label{fig5}
\end{figure}

\placefigure{fig5}     % Figure 5 - Mean spectrum of HD 54662 

\begin{figure}
\begin{center}
{\includegraphics[angle=90,height=12cm]{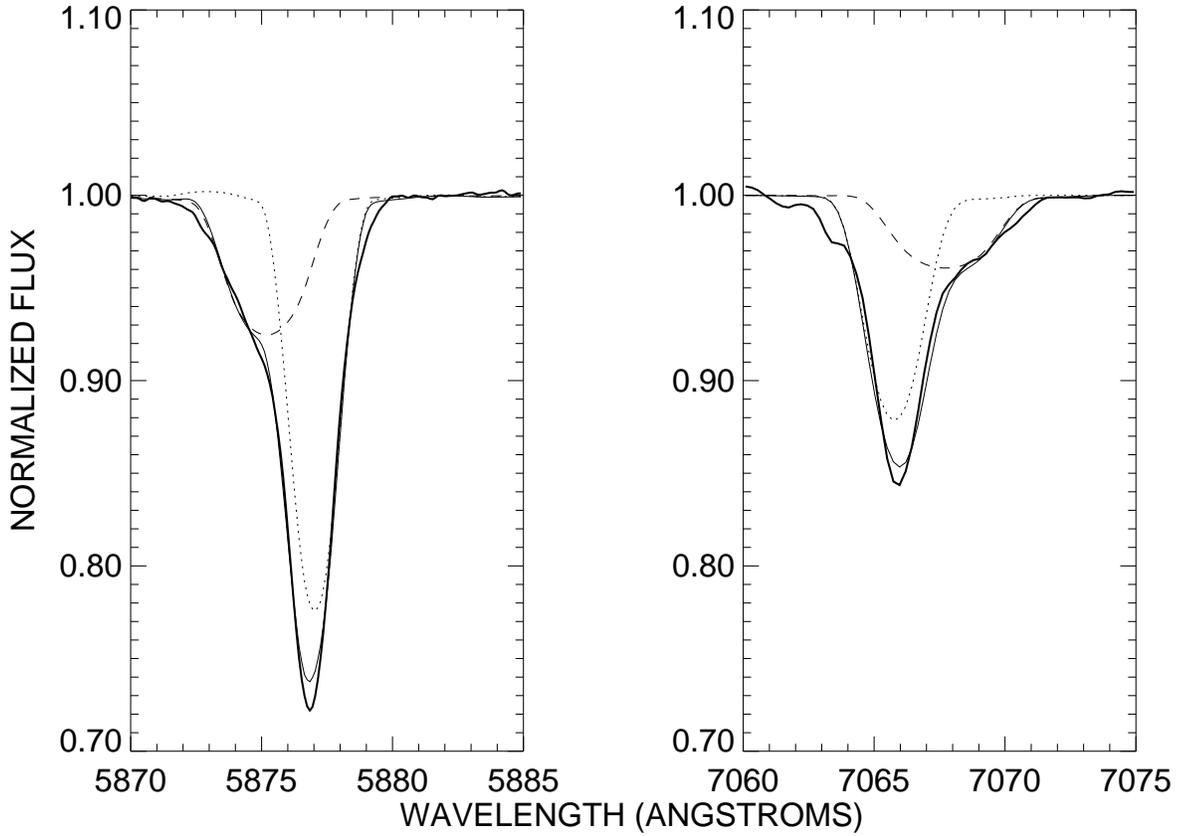}}
\end{center}
\caption{The mean line profiles of HD~54662 for \ion{He}{1} $\lambda 5876$
({\it left}; \citealt{ful90}) and \ion{He}{1} $\lambda 7065$ ({\it right}; 
this work). The observations are plotted as {\it thick lines}, the 
combined model fit as {\it thin lines}, and the individual component  
profiles are represented as {\it dotted/dashed lines} for the
primary and secondary, respectively.}
\label{fig6}
\end{figure}

\placefigure{fig6}     % Figure 6 - Composite profiles of HD 54662 for both He lines, modshift1.pro written 9-21-06

We made preliminary two component fits of the 
blended \ion{He}{1} lines ($\lambda 5876$ for the 
spectra obtained by \citealt{ful90} and  
$\lambda 7065$ for this work) using
TLUSTY/SYNSPEC models.  We used the temperature and 
gravity calibrations of 
\citet{mar05} to select parameters for the composite 
model profiles to 
fit our observations.  Our model spectra for the primary star  
are based upon an assumed type of O6.5~V \citep{wal72}. 
We constructed model spectra for the secondary 
for spectral subtypes of O7~V -- O9.5~V.   
Next, we compared our 
observed mean line profiles to  
these models applying the appropriate 
flux ratio (from $\Delta M_{V}$ 
in \citealt{mar05}) for each 
spectral component in the shifted, combined line profiles.  
In each trial for a given secondary spectral type, the 
only variables were the component radial velocities
and the secondary's projected rotational velocity 
$V_2 \sin i$ (we assumed
$V_1 \sin i = 70$~\kms;~\citealt{con77}).  
Our best match for the 
secondary was made with an O9~V 
subtype and $V_2 \sin i = 110 \pm 10$ \kms, 
which yields a flux ratio of $F_{2}/F_{1} = 
0.51$.  Our fits of \ion{He}{1} $\lambda7065$ 
required us to make small and equal adjustments to the model 
line depths.  The resulting fits are shown in
Figure~6.  We caution that an uncertainty in 
$V_2 \sin i$ has a large effect upon the best 
fit line shifts and flux ratio results. 

The wavelength shifts made to fit these 
composite line profiles 
provide us with average velocities for the 
primary and secondary components for each observing run.  
Assuming that the true anomaly $\nu$ and the longitude 
of periastron $\omega$ are known from the preliminary
orbital fit (Table~4), we may estimate 
the systematic velocity $\gamma$ and 
semiamplitude $K$ by making a least-squares, linear fit
of these three velocities using 
\begin{equation}
V_{r}= \gamma_{1,2} \pm K_{1,2} [\cos(\nu+\omega)+e \cos\omega].
\end{equation}
This solution gives semiamplitudes of $K_{1} = 29\pm4$~\kms~and 
$K_{2} = 75\pm7$~\kms~and systemic velocities of 
$\gamma_{1} = 45\pm3$~\kms~and $\gamma_{2} = 40\pm6$~\kms.
This estimate of the
secondary radial velocity curve also allows us to 
compute the component minimum masses of the system,
$M_1\sin ^{3}i\approx 41.5 \pm 7.6$~$M_\odot$ and 
$M_2\sin ^{3}i\approx 16.0 \pm 3.4$~$M_\odot$. 
The radial velocity curves 
for these solutions for the primary 
({\it dashed line}) and secondary 
({\it dot-dashed line}) are also 
plotted in Figure~2, along with the 
time-averaged radial velocities from the 
two-component fits.  This analysis of the line blending
problem clearly illustrates how the presence of the 
blended secondary spectrum skews the velocity 
measurements for the primary (Table~3) towards 
the system's center of mass, resulting in a 
semiamplitude (Table~4) that is approximately a factor 
of two smaller than the actual value.

%%%%%%%%%%%%%%%%%%%%%%%%%%%%%%%%%%%%%%%%%%%%%%%%%%%%%%%%

\section{Discussion}                        % Section 6
    
One of the motivations for this study
was to find long period binaries that 
may be resolved by optical 
long baseline interferometry.  The CHARA Array,
for example, can resolve binaries with 
angular separations as small as 1 mas
\citep{ten05}.  To determine the 
angular separation of the binaries 
components, we re-estimated their distances by 
fitting their observed 
spectral energy distribution (SED) with 
a model SED to find the angular stellar 
diameters that we then compared with stellar
radii estimates for their spectral classifications.  
For each binary, the 
model temperatures, gravities, and flux ratios were
applied to create a combined model flux distribution over 
a range of $1200-30000$ \AA. The galactic 
extinction curve from \citet{fit99} was then 
applied to the model SED to fit the observed 
photometry for each target.  The 
observed SED includes ultraviolet fluxes 
({\it IUE}; TD-1, \citealt{tho78}) 
and $UBV$ \citep*{nek80}, 
$uvby$ \citep{hau98}, 
and {\it 2MASS} $JHK$ infrared magnitudes \citep{skr06,cut03},
all of which were transformed into calibrated flux
measurements \citep*{col96,gra98,coh03}.  
The best fit parameters for reddening 
$E(B-V)$,  ratio of total-to-selective
extinction $R$, and the limb darkened angular 
diameter for the primary $\theta_{LD}$ (from 
the flux normalization) are listed in Table~6.
Figures 7 and 8 show the SED plots of these best 
fits for HD~37366 and HD~54662, respectively.

\begin{figure}
\begin{center}
{\includegraphics[angle=90,height=12cm]{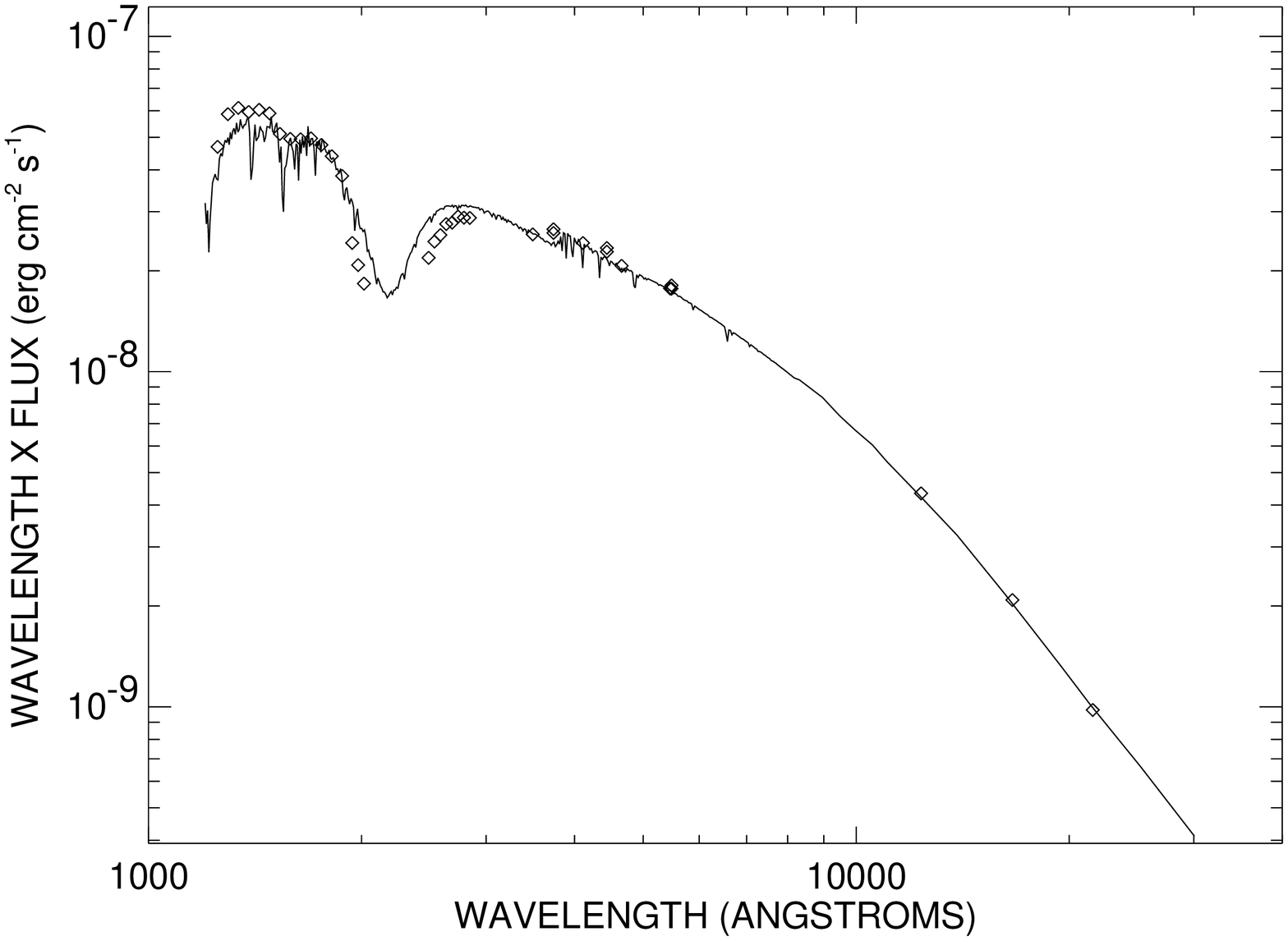}}
\end{center}
\caption{Best fit SED for HD~37366.  The {\it solid line} indicates
the combined model flux for the binary, and the {\it diamonds}
represent the photometric observations (described in the text).}
\label{fig7}
\end{figure}

\begin{figure}
\begin{center}
{\includegraphics[angle=90,height=12cm]{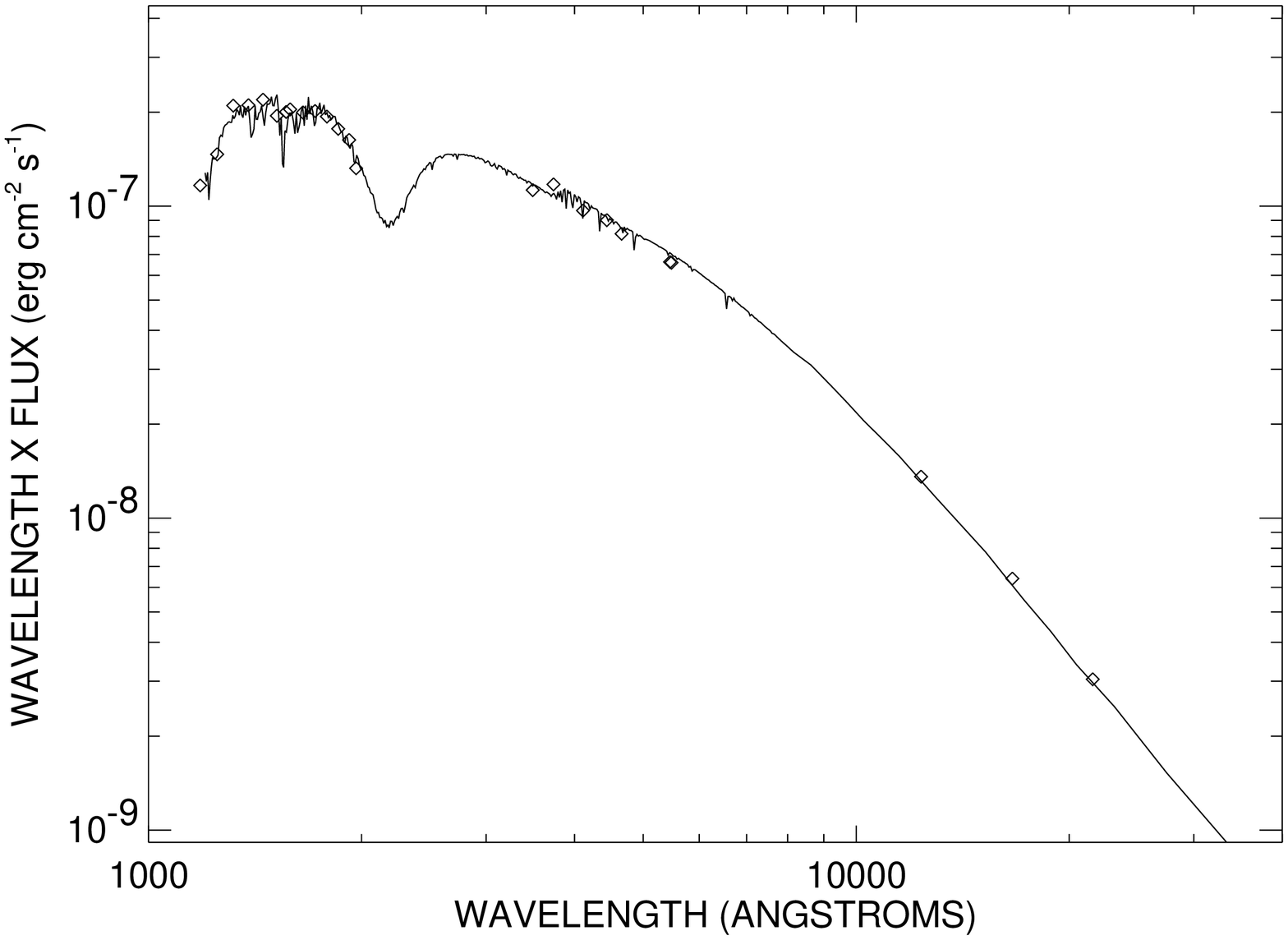}}
\end{center}
\caption{Best fit SED for HD~54662.  The {\it solid line} indicates
the combined model flux for the binary, and the {\it diamonds}
represent the photometric observations (described in the text).}
\label{fig8}
\end{figure}

\placetable{tab6}	% Table 6 - SED parameters 
\placefigure{fig7}     % Figure 7 - SED plot for HD37366 
\placefigure{fig8}     % Figure 8 - SED plot for HD54662

We then compared the expected theoretical 
radii for the primary stars, 
based upon their spectral classifications 
as main sequence stars \citep{mar05}, with the angular 
sizes to obtain distances of 1.38 and 1.23~kpc
for HD~37366 and HD~54662, respectively (Table~6).
These estimates are consistent with the accepted 
distances to their home associations \citep{hum78}.
The distance ranges given in Table~6 reflect 
the change in stellar radius between ZAMS 
luminosity class V and giant 
luminosity class III.  The mid-point increase in 
size is $\triangle R \approx 3 R_\odot$ for our stars, 
and we adopted this difference to estimate the 
associated range in distance.  Errors will also 
result from the spread in radius for each 
spectral sub-type bin, but this is 
quite small ($\pm 0.4 R_\odot$) compared to 
the luminosity range. The error in  
$\theta_{LD}$ from the SED fit contributes 
only $\approx 1\%$ in the distance error budget.
  
The binary semimajor axis $a$ 
was found using Kepler's Third Law,
the derived orbital period, and the stellar 
mass calibrations from \citet{mar05} 
(for O stars) and \citet{har88} (for B stars).  
The results for the maximum angular separation 
$\rho_{\rm max}$ for the projected 
elliptical orbit are also presented in Table~6
where we give the range in $\rho_{\rm max}$ 
associated with the range in distance.     
These separations are too small for 
speckle resolution ($\rho > 0\farcs035$ for 
$\triangle m <3.0$), but they are close to 
or above the limits of 
long baseline interferometry.  The HD~54662
binary system in particular may prove to be 
important target for mass determination
by interferometry.

%%%%%%%%%%%%%%%%%%%%%%%%%%%%%%%%%%%%%%%%%%%%%%%%%%%%%%%%%%%%%%%

\acknowledgments

We thank Daryl Willmarth, Paul Rybski, Eric Phillips,
Travis Fischer, and the KPNO staff for their assistance
in making these observations possible.
This material is based on work supported by the
National Science Foundation under Grants 
AST-0506573 and AST-0606861.
Bolton's research is partially supported by 
a Natural Sciences and Engineering Research 
Council of Canada (NSERC) Discovery Grant.
Fullerton held an NSERC Post-graduate Scholarship 
during the course of this work.
Some of the data presented in this paper were 
obtained from the Multimission Archive at the 
Space Telescope Science Institute (MAST). 
STScI is operated by the Association of 
Universities for Research in Astronomy, 
Inc., under NASA contract NAS5-26555. 
Support for MAST for non-HST data is provided 
by the NASA Office of Space Science via grant 
NAG5-7584 and by other grants and contracts.
This publication makes use of data products 
from the Two Micron All Sky Survey, which is 
a joint project of the University of Massachusetts 
and the Infrared Processing and Analysis 
Center/California Institute of Technology, 
funded by the National Aeronautics and Space 
Administration and the National Science Foundation.

%%%%%%%%%%%%%%%%%%%%%%%%%%%%%%%%%%%%%%%%%%%%%%%%%%%%%%%%%%%%%%%

% The bibliography starts here.

\bibliographystyle{apj}             % Please learn to use the
                                    % formatting of Latex's Bibtex. It
                                    % will make your life easier.
% apj.bst should be in this directory as well as apj-jour.bib and reference paper.bib
\bibliography{apj-jour,paper}       % "paper.bib" contains all my
                                    % references. "apj-jour.bib"
                                    % contains abbreviations of
                                    % journals.

%%%%%%%%%%%%%%%%%%%%%%%%%%%%%%%%%%%%%%%%%%%%%%%%%%%%%%%%%%%%%%%
% Tables

\newpage
% Table 1
\begin{deluxetable}{lcccccccc}
\tabletypesize{\scriptsize}
\tablewidth{0pt}
\tablenum{1}
\tablecaption{HD~37366 Radial Velocity Measurements\label{tab1}}
\tablehead{
\colhead{HJD}	        &
\colhead{Telescope/}	&
\colhead{Orbital}	&
\colhead{$V_1$}	        &
\colhead{$\sigma_{1}$}	&
\colhead{$(O-C)_1$}	&
\colhead{$V_2$}		&
\colhead{$\sigma_{2}$}	&
\colhead{$(O-C)_2$}	\\
\colhead{($-$2,400,000)}	&
\colhead{Band}	&
\colhead{Phase}	 &
\colhead{(km s$^{-1}$)} &
\colhead{(km s$^{-1}$)} &
\colhead{(km s$^{-1}$)} &
\colhead{(km s$^{-1}$)} &
\colhead{(km s$^{-1}$)} &
\colhead{(km s$^{-1}$)} }
\startdata
46821.612  & IUE/UV &  0.303 & 
\phn\phs     $  76.9$ & 5.0 &\phn\phs $   0.7$ & \phn   $ -57.3$ & 10.0 & \phn\phs $   4.3$\\
46866.116  & IUE/UV &  0.702 & 
\phn\phs     $  13.2$ & 5.0 &\phn     $  -0.5$ & \phn \nodata & \nodata & \phs \nodata\\
51817.934  & CF/Red & 0.327 & 
\phn\phs     $  75.9$ & 1.3 &\phn     $  -1.3$ & \phn    $ -64.5$ & \phn 2.2 &\phn     $  -1.5$ \\
51818.938  & CF/Red & 0.358 & 
\phn\phs     $  76.2$ & 1.3 &\phn     $  -1.0$ & \phn    $ -67.1$ &\phn2.2 &\phn     $  -4.1$ \\
51819.929  & CF/Red & 0.389 & 
\phn\phs     $  75.5$ & 1.3 &\phn     $  -0.6$ & \phn    $ -53.2$ &\phn2.2 &\phn\phs $   8.3$ \\
51820.922  & CF/Red & 0.420 & 
\phn\phs     $  74.7$ & 1.3 &\phn\phs $   0.8$ & \phn    $ -63.8$ &\phn2.2 &\phn     $  -5.2$ \\
51821.918  & CF/Red & 0.452 & 
\phn\phs     $  71.2$ & 1.3 &\phn\phs $   0.4$ & \phn    $ -51.9$ &\phn2.2 &\phn\phs $   2.5$ \\
51822.918  & CF/Red & 0.483 & 
\phn\phs     $  64.1$ & 1.3 &\phn     $  -2.7$ & \phn    $ -49.4$ &\phn2.2 &\phn     $  -0.3$ \\
51823.853  & CF/Red & 0.513 & 
\phn\phs     $  62.2$ & 1.3 &\phn     $  -0.1$ & \phn    $ -38.1$ &\phn2.2 &\phn\phs $   4.9$ \\
51823.980  & CF/Red & 0.517 & 
\phn\phs     $  60.1$ & 1.3 &\phn     $  -1.5$ & \phn    $ -41.7$ &\phn2.2 &\phn\phs $   0.4$ \\
51824.881  & CF/Red & 0.545 & 
\phn\phs     $  55.8$ & 1.3 &\phn     $  -0.7$ & \phn    $ -38.8$ &\phn2.2 &\phn     $  -3.6$ \\
51824.997  & CF/Red & 0.549 & 
\phn\phs     $  57.1$ & 1.3 &\phn\phs $   1.3$ & \phn    $ -36.9$ &\phn2.2 &\phn     $  -2.6$ \\
51830.904  & CF/Red & 0.734 & 
\phn\phn\phs $   0.2$ & 1.3 &\phn     $  -1.1$ & \phn\phs$  45.9$ &\phn2.2 &\phn\phs $   7.4$ \\
51889.881  & CF/Red & 0.588 & 
\phn\phs     $  47.9$ & 1.3 &\phn\phs $   0.5$ & \phn    $ -26.0$ &\phn2.2 &\phn     $  -3.0$ \\
51890.819  & CF/Red & 0.617 & 
\phn\phs     $  42.0$ & 1.3 &\phn\phs $   2.0$ & \phn    $ -20.9$ &\phn2.2 &\phn     $  -7.6$ \\
51892.787  & CF/Red & 0.679 & 
\phn\phs     $  20.9$ & 1.3 &\phn     $  -0.8$ & \phn\phs$  17.9$ &\phn2.2 &\phn\phs $   6.7$ \\
51893.855  & CF/Red & 0.713 & 
\phn\phn\phs $   9.6$ & 1.3 &\phn     $  -0.2$ & \phn\phs$  26.5$ &\phn2.2 &\phn     $  -0.7$ \\
51894.780  & CF/Red & 0.742 & 
\phn\phn     $  -1.6$ & 1.3 &\phn\phs $   0.3$ & \phn\phs$  53.6$ &\phn2.2 &\phs     $  10.9$ \\
51894.856  & CF/Red & 0.744 & 
\phn\phn     $  -3.3$ & 1.3 &\phn     $  -0.5$ & \phn\phs$  47.2$ &\phn2.2 &\phn\phs $   3.2$ \\
51895.874  & CF/Red & 0.776 & 
\phn         $ -19.6$ & 1.3 &\phn     $  -2.4$ & \phn\phs$  58.7$ &\phn2.2 &\phn     $  -4.5$ \\
51896.802  & CF/Red & 0.805 & 
\phn         $ -32.7$ & 1.3 &\phn     $  -1.0$ & \phn\phs$  78.4$ &\phn2.2 &\phn     $  -4.2$ \\
51896.914  & CF/Red & 0.809 & 
\phn         $ -34.6$ & 1.3 &\phn     $  -1.0$ & \phn\phs$  78.8$ &\phn2.2 &\phn     $  -6.1$ \\
51897.802  & CF/Red & 0.837 & 
\phn         $ -46.9$ & 1.3 &\phn\phs $   1.7$ & \phs    $ 100.7$ &\phn2.2 &\phn     $  -4.4$ \\
51897.910  & CF/Red & 0.840 & 
\phn         $ -51.4$ & 1.3 &\phn     $  -0.8$ & \phn\phs$  96.0$ &\phn2.2 &         $ -11.6$ \\
51898.811  & CF/Red & 0.868 & 
\phn         $ -66.2$ & 1.3 &\phn\phs $   0.3$ & \phs    $ 128.1$ &\phn2.2 &\phn     $  -0.8$ \\
51898.922  & CF/Red & 0.872 & 
\phn         $ -68.7$ & 1.3 &\phn     $  -0.2$ & \phs    $ 133.2$ &\phn2.2 &\phn\phs $   1.6$ \\
51899.809  & CF/Red & 0.900 & 
\phn         $ -83.5$ & 1.3 &\phn     $  -0.1$ & \phs    $ 153.0$ &\phn2.2 &\phn\phs $   1.7$ \\
51899.915  & CF/Red & 0.903 & 
\phn         $ -84.4$ & 1.3 &\phn\phs $   0.5$ & \phs    $ 154.2$ &\phn2.2 &\phn\phs $   0.7$ \\
51900.802  & CF/Red & 0.931 & 
\phn         $ -95.5$ & 1.3 &\phn\phs $   0.6$ & \phs    $ 171.8$ &\phn2.2 &\phn\phs $   3.5$ \\
51900.908  & CF/Red & 0.934 & 
\phn         $ -96.3$ & 1.3 &\phn\phs $   0.7$ & \phs    $ 172.6$ &\phn2.2 &\phn\phs $   3.1$ \\
51901.788  & CF/Red & 0.962 & 
\phn         $ -99.8$ & 1.3 &\phn\phs $   0.3$ & \phs    $ 173.7$ &\phn2.2 &\phn\phs $   0.2$ \\
51901.917  & CF/Red & 0.966 & 
             $-100.5$ & 1.3 &\phn     $  -0.7$ & \phs    $ 177.2$ &\phn2.2 &\phn\phs $   4.1$ \\
53291.928  & CF/Red & 0.651 & 
\phn\phs     $  30.8$ & 2.6 & \phn\phs $   0.2$ &  \phn \nodata & \phn\nodata &  \phs \nodata \\
53292.984  & CF/Red & 0.684 & 
\phn\phs     $  20.6$ & 0.4 & \phn\phs $   0.6$  & \phn \nodata & \phn\nodata &  \phs \nodata  \\
53658.997\tablenotemark{a}  & 2.1m/Blue & 0.187 & 
\phn\phs     $  52.6$ & 6.7 & \phn     $  -0.8$  & \phn \nodata & \phn\nodata & \phs \nodata  \\
53659.000\tablenotemark{a}  & 2.1m/Blue & 0.187 & 
\phn\phs     $  51.5$ & 1.4 & \phn     $  -2.0$  & \phn \nodata & \phn\nodata & \phs \nodata  \\
53663.010\tablenotemark{a}  & 2.1m/Blue & 0.313 & 
\phn\phs     $  71.6$ & 0.4 & \phn     $  -5.0$  & \phn \nodata & \phn\nodata & \phs \nodata  \\
53663.989\tablenotemark{a}  & 2.1m/Blue & 0.344 & 
\phn\phs     $  89.0$ & 4.2 & \phs     $  11.7$  & \phn \nodata & \phn\nodata & \phs \nodata  \\
53684.903  & CF/Blue  & 0.001 & 
\phn         $ -85.7$ & 1.3 &\phn\phs $   0.8$ & \phs     $ 155.4$ & \phn4.2 &\phn\phs $   0.1$ \\
53686.902  & CF/Blue  & 0.064 & 
\phn         $ -30.1$ & 0.7 &\phn\phs $   0.9$ & \phn\phs $  76.3$ & \phn9.8 &\phn     $  -4.9$ \\
53688.846  & CF/Blue  & 0.125 & 
\phn\phs     $  21.9$ & 0.9 &\phn\phs $   0.9$ & \phn\phs $  13.1$ & \phn5.2 &\phn\phs $   1.3$ \\
54020.972  & CF/Red  & 0.564 &
\phn\phs     $  51.2$ & 1.8 & \phn     $  -1.1$ & \phn $ -25.8$ & \phn2.2 & \phn\phs $   4.1$ \\
54024.926  & CF/Red  & 0.688 &
\phn\phs     $  17.4$ & 2.8 & \phn     $  -1.0$ & \phn\phs  $  13.9$ & \phn3.4 &\phn   $  -1.1$ \\
54030.021  & CF/Blue  & 0.848 &
\phn         $ -56.0$ & 2.0 & \phn  $  -0.8$ & \phs     $ 103.2$ & \phn2.4 & \phn $  -9.3$\\
54031.999  & CF/Blue  & 0.911 &
\phn         $ -90.1$ & 2.9 & \phn     $  -1.7$ & \phs  $ 144.4$ & \phn3.5 &   $ -12.0$  \\
\enddata
%\caption{Radial velocities and observations for HD37366. All measurements were made 
%at the KPNO Coud\'{e} Feed telescope during October and December of 2000, unless 
%otherwise indicated.}
\tablenotetext{a}{zero weight}
%\tablenotetext{a}{{\it IUE}, 1987}
%\tablenotetext{b}{KPNO Coud\'{e} Feed, 2004} 
%\tablenotetext{c}{KPNO 2.1m, 2005}
%\tablenotetext{d}{KPNO Coud\'{e} Feed, 2005}
%\tablenotetext{e}{KPNO Coud\'{e} Feed, 2006}
\end{deluxetable}
\newpage

%Table 2
\begin{deluxetable}{lccc}
\tabletypesize{\scriptsize}
\tablewidth{0pc}
\tablenum{2}
\tablecaption{Orbital Elements for HD 37366\label{tab2}}
\tablehead{
\colhead{Element}	& 
\colhead{Primary}	&
\colhead{Secondary}	&
\colhead{Joint Solution}	}
\startdata
$P$~(days)                 \dotfill & $31.8187 \pm 0.0004$\phn	& $31.822 \pm 0.002$\phn & $31.8188$\tablenotemark{a} \\
$T_1$ (HJD--2,400,000)     \dotfill & $53653.013\pm0.04$\phn\phn\phn\phn\phn & \nodata & $53653.02$\tablenotemark{a} \\
$T_2$ (HJD--2,400,000)     \dotfill & \nodata & $53653.15 \pm0.19$\phn\phn\phn\phn & $53653.02$\tablenotemark{a} \\
$e_1$                      \dotfill & $0.329\pm0.003$ & \nodata & $0.330$\tablenotemark{a}	\\
$e_2$                      \dotfill & \nodata         & \phn$0.35 \pm0.012$ & $0.330$\tablenotemark{a}	\\
$\omega_1$ (deg)           \dotfill & $211.4 \pm0.6$\phn\phn	& \nodata   & $211.6$\tablenotemark{a}  \\
$\omega_2$ (deg)           \dotfill & \nodata	      & $212 \pm2$\phn	    & $211.6$\tablenotemark{a}\\
$K_1$ (km s$^{-1}$)        \dotfill & $88.6 \pm0.3$\phn & \nodata           & $88.7 \pm0.2$\phn  \\
$K_2$ (km s$^{-1}$)        \dotfill & \nodata	      & $118.4 \pm1.6$\phn\phn & $117.4\pm1.2$\phn\phn  \\
$\gamma_1$ (km s$^{-1}$)   \dotfill & $13.3 \pm0.2$\phn & \nodata           & $13.3\pm0.2$\phn  \\
$\gamma_2$ (km s$^{-1}$)   \dotfill & \nodata	      & $20.6 \pm1.3$\phn   & $21.6\pm0.9$\phn \\
$M_1\sin ^{3}i$ ($M_\odot$)\dotfill & $13.8 \pm 0.3$\phn & \nodata          & $13.9 \pm0.3$\phn  \\
$M_2\sin ^{3}i$ ($M_\odot$)\dotfill & \nodata	      &	$10.5 \pm0.1$       & $10.42 \pm0.08$\phn  \\
$a_1 \sin i$ ($R_\odot$)   \dotfill & $52.6 \pm0.2$\phn	& \nodata	    & $ 52.62\pm0.13$\phn  \\
$a_2 \sin i$ ($R_\odot$)   \dotfill & \nodata         & $69.7 \pm1.0$\phn   & $69.7 \pm0.7$\phn  \\
$\sigma_1$ (km s$^{-1}$)   \dotfill & \phn$1.1$       & \nodata             & \phn $1.0$ \\
$\sigma_2$ (km s$^{-1}$)   \dotfill & \nodata         & \phn$5.3$           & \phn$5.4$  \\
\enddata
\tablenotetext{a}{Fixed.}
\end{deluxetable}

\newpage

% Table 3
\begin{deluxetable}{lccccc}
\tabletypesize{\scriptsize}
\tablewidth{0pt}
\tablenum{3}
\tablecaption{HD~54662 Primary Radial Velocity Measurements\label{tab3}}
\tablehead{
\colhead{HJD}   &
\colhead{Telescope/}	&
\colhead{Orbital}       &
\colhead{$V_{r}$}	&
\colhead{$\sigma$(line$-$line)}	&
\colhead{$O-C$}      \\
\colhead{($-$2,400,000)}&
\colhead{Band}		&
\colhead{Phase}         &
\colhead{(km s$^{-1}$)} &
\colhead{(km s$^{-1}$)} &
\colhead{(km s$^{-1}$)} }
\startdata
46426.830 & CFHT/Yellow &  0.191 &
\phn\phs     $  62.2$ & 2.0 & \phn     $  -0.3$ \\
46426.904 & CFHT/Yellow &  0.191 &
\phn\phs     $  61.3$ & 1.6 & \phn     $  -1.1$ \\
46426.980 & CFHT/Yellow &  0.191 &
\phn\phs     $  61.8$ & 1.6 & \phn     $  -0.6$ \\
46427.802 & CFHT/Yellow & 0.192 &
\phn\phs     $  61.3$ & 1.7 & \phn     $  -1.2$ \\
46427.869 & CFHT/Yellow &  0.192 &
\phn\phs     $  61.6$ & 2.2 & \phn     $  -0.9$ \\
46427.925 & CFHT/Yellow &  0.193 &
\phn\phs     $  61.1$ & 2.0 & \phn     $  -1.4$ \\
46427.973 & CFHT/Yellow &  0.193 &
\phn\phs     $  61.5$ & 1.6 & \phn     $  -1.0$ \\
46428.036 & CFHT/Yellow &  0.193 &
\phn\phs     $  61.6$ & 1.4 & \phn     $  -0.9$ \\
46428.132 & CFHT/Yellow &  0.193 &
\phn\phs     $  62.0$ & 2.1 & \phn     $  -0.5$ \\
46428.805 & CFHT/Yellow &  0.194 &
\phn\phs     $  62.2$ & 2.0 & \phn     $  -0.3$ \\
46429.021 & CFHT/Yellow &  0.194 &
\phn\phs     $  61.4$ & 1.9 & \phn     $  -1.2$ \\
46429.814 & CFHT/Yellow &  0.196 &
\phn\phs     $  61.8$ & 1.7 & \phn     $  -0.8$ \\
46429.883 & CFHT/Yellow &  0.196 &
\phn\phs     $  61.5$ & 2.1 & \phn     $  -1.1$ \\
46432.853 & CFHT/Yellow &  0.201 &
\phn\phs     $  61.4$ & 2.1 & \phn     $  -1.3$ \\
46432.897 & CFHT/Yellow &  0.201 &
\phn\phs     $  60.9$ & 1.4 & \phn     $  -1.8$ \\
46432.999 & CFHT/Yellow &  0.202 &
\phn\phs     $  62.2$ & 2.0 & \phn     $  -0.5$ \\
46433.093 & CFHT/Yellow &  0.202 &
\phn\phs     $  61.2$ & 1.5 & \phn     $  -1.6$ \\
51817.967 & CF/Red  &  0.855 &
\phn\phs     $  33.2$ & 0.6 & \phn\phs $   0.0$ \\
51818.962 & CF/Red  &  0.857 &
\phn\phs     $  33.5$ & 2.5 & \phn\phs $   0.4$ \\
51819.962 & CF/Red  &  0.859 &
\phn\phs     $  33.7$ & 1.1 &\phn\phs $   0.7$ \\
51820.990 & CF/Red  &  0.860 &
\phn\phs     $  33.0$ & 2.1 & \phn\phs $   0.1$ \\
51821.968 & CF/Red  &  0.862 &
\phn\phs     $  33.0$ & 0.9 & \phn\phs $   0.2$ \\
51822.941 & CF/Red  &  0.864 &
\phn\phs     $  34.6$ & 3.7 & \phn\phs $   1.9$ \\
51823.957 & CF/Red  &  0.866 &
\phn\phs     $  34.5$ & 3.4 & \phn\phs $   1.8$ \\
51824.903 & CF/Red  &  0.867 &
\phn\phs     $  35.2$ & 1.2 & \phn\phs $   2.6$ \\
51889.990 & CF/Red  &  0.984 &
\phn\phs     $  37.5$ & 2.6 & \phn\phs $   0.7$ \\
51890.923 & CF/Red  &  0.986 &
\phn\phs     $  37.1$ & 1.2 & \phn\phs $   0.0$ \\
51892.899 & CF/Red  &  0.989 &
\phn\phs     $  35.0$ & 2.0 & \phn     $  -2.6$ \\
51893.926 & CF/Red  &  0.991 &
\phn\phs     $  37.5$ & 1.9 & \phn     $  -0.3$ \\
51894.882 & CF/Red  &  0.993 &
\phn\phs     $  37.5$ & 1.5 & \phn     $  -0.6$ \\
51894.956 & CF/Red  &  0.993 &
\phn\phs     $  39.0$ & 2.3 & \phn\phs $   0.9$ \\
51895.934 & CF/Red  &  0.995 &
\phn\phs     $  37.6$ & 1.3 & \phn     $  -0.8$ \\
51896.033 & CF/Red  &  0.995 &
\phn\phs     $  39.0$ & 1.6 & \phn\phs $   0.6$ \\
51896.881 & CF/Red  &  0.996 &
\phn\phs     $  39.4$ & 0.6 & \phn\phs $   0.8$ \\
51896.952 & CF/Red  &  0.997 &
\phn\phs     $  37.0$ & 1.6 &\phn     $  -1.7$ \\
51897.879 & CF/Red  &  0.998 &
\phn\phs     $  37.9$ & 1.7 & \phn     $  -1.0$ \\
51897.943 & CF/Red  &  0.998 &
\phn\phs     $  37.1$ & 1.6 & \phn     $  -1.8$ \\
51898.891 & CF/Red  &  0.000 &
\phn\phs     $  39.1$ & 2.5 & \phn     $  -0.1$ \\
51898.953 & CF/Red  &  0.000 &
\phn\phs     $  38.3$ & 2.0 & \phn     $  -0.9$ \\
51899.885 & CF/Red  &  0.002 &
\phn\phs     $  37.4$ & 0.8 & \phn     $  -2.1$ \\
51899.947 & CF/Red  &  0.002 &
\phn\phs     $  38.3$ & 0.5 & \phn     $  -1.2$ \\
51900.878 & CF/Red  &  0.004 &
\phn\phs     $  39.0$ & 2.9 & \phn     $  -0.8$ \\
51900.940 & CF/Red  &  0.004 &
\phn\phs     $  38.7$ & 3.1 & \phn     $  -1.1$ \\
51901.885 & CF/Red  &  0.005 &
\phn\phs     $  39.5$ & 2.9 & \phn     $  -0.6$ \\
51901.949 & CF/Red  &  0.006 &
\phn\phs     $  39.1$ & 2.8 & \phn     $  -1.0$ \\
54020.025 & CF/Red &  0.802 &
\phn\phs     $  33.9$ & 7.5 & \phn     $  -2.5$ \\
54024.964 & CF/Red &  0.811 &
\phn\phs     $  31.8$ & 5.6 & \phn     $  -4.0$ \\
54027.025 & CF/Blue &  0.815 &
\phn\phs     $  35.8$ & 5.0 & \phn\phs $   0.2$ \\
54028.964 & CF/Blue &  0.819 &
\phn\phs     $  34.4$ & 6.4 & \phn     $  -0.9$ \\
54030.961 & CF/Blue &  0.822 &
\phn\phs     $  37.4$ & 4.4 & \phn\phs $   2.3$ \\
54032.012 & CF/Blue &  0.824 &
\phn\phs     $  35.2$ & 4.3 & \phn\phs $   0.2$ \\
\enddata
%\tablenotetext{a}{\citet{ful90}}
%\tablenotetext{b}{KPNO Coud\'{e} Feed, 2006}
\end{deluxetable}

\newpage

%Table 4: edited 11-9-06 for solutions with cf 2006 Vr's 
\begin{deluxetable}{lc}
%\tabletypesize{\scriptsize}
\tablewidth{0pc}
\tablenum{4}
\tablecaption{Preliminary Orbital Elements for HD 54662\label{tab4}}
\tablehead{
\colhead{Element} &
\colhead{Value}}
\startdata
$P$~(days)              \dotfill    & $557.8 \pm 0.3 \phn\phn $          \\
$T$ (HJD--2,400,000)    \dotfill    & $22333 \pm  5\phn\phn\phn\phn$        \\
$e$                     \dotfill    & $0.28 \pm 0.04 $           \\
$\omega$ (deg)          \dotfill    & $238 \pm 5 \phn\phn $              \\
$K$\tablenotemark{a} (km s$^{-1}$) \dotfill & $15.9 \pm  0.5 \phn$           \\
$\gamma$ (km s$^{-1}$)  \dotfill    & $49.9 \pm  0.6 \phn$      \\
$f(m)$\tablenotemark{a}  ($M_\odot$)     \dotfill    & $0.20  \pm 0.02$       \\
$a_1\sin i$\tablenotemark{a} ($R_\odot$) \dotfill    & $168 \pm 6 \phn\phn$             \\
rms (km s$^{-1}$)       \dotfill    & $\phn 3.3$                     \\
\enddata
\smallskip
\tablenotetext{a}{\ Lower limit due to line blending.}
\end{deluxetable}
\newpage

%Table 5
\begin{deluxetable}{lc}
%\tabletypesize{\scriptsize}
\tablewidth{0pc}
\tablenum{5}
\tablecaption{Stellar Parameters for HD~37366\label{tab5}}
\tablehead{
\colhead{Parameter} &
\colhead{Value} }
\startdata
$V_1\sin i$ (\kms)	&	$30 \pm 10$ 	\\
$V_2\sin i$ (\kms)	&	$100 \pm 10$ 	\\
$T_{\rm eff,1}$ (kK)	&	$33 \pm 1$ 	\\
$T_{\rm eff,2}$ (kK)	&	$30 \pm 1$ 	\\
$\log g_{1}$ (cgs)	&	$4.0 \pm 0.1$		\\
$\log g_{2}$ (cgs)	&	$4.5 \pm 0.2$		\\	
$F_{2}/F_{1}$		&	$0.35 \pm 0.05$	\\
$\Delta M_{V}$		&	$1.1 \pm 0.1$		\\
\enddata
\end{deluxetable}

%Table 6
\begin{deluxetable}{lcc}
%\tabletypesize{\scriptsize}
\tablewidth{0pc}
\tablenum{6}
\tablecaption{SED Parameters\label{tab6}}
\tablehead{
\colhead{Parameter} &
\colhead{HD~37366} &
\colhead{HD~54662}}
\startdata
Primary Type          \dotfill & \phn O9.5~V\tablenotemark{a} & O6.5~V\tablenotemark{a} \\
Secondary Type        \dotfill &  B0$-$1~V	              &	O9~V\phn   \\
$E(B-V)$(mag)         \dotfill & $0.39 \pm 0.01$              & $0.32 \pm 0.01$ \\
$R_V$(mag)	      \dotfill & $3.59 \pm 0.01$	      & $2.82 \pm 0.01$ \\
$\theta_{LD}$($\mu$as)\tablenotemark{b} \dotfill & $48.4 \pm 3.0\phn$	      & $72.7\pm3.4\phn$ \\
$d$ (kpc)             \dotfill & $1.38-1.92$                  & $1.23-1.53$ \\		
$\rho_{\rm max}$ (mas)          \dotfill & $0.4-0.5$	              & $3.7-4.7$ \\
\enddata
\tablenotetext{a}{\citet{wal72,wal73}}
\tablenotetext{b}{Primary}
\end{deluxetable}

%%%%%%%%%%%%%%%%%%%%%%%%%%%%%%%%%%%%%%%%%%%%%%%%%%%%%%%%%%%%%%

% Figure captions

\clearpage

\input{epsf}
% Figure 1

\clearpage

\input{epsf}
% Figure 2

\clearpage

\input{epsf}
% Figure 3

\clearpage

\input{epsf}
% Figure 4

\clearpage

\input{epsf}
% Figure 5

\clearpage

\input{epsf}
% Figure 6

\input{epsf}
% Figure 7

\input{epsf}
% Figure 8

%%%%%%%%%%%%%%%%%%%%%%%%%%%%%%%%%%%%%%%%%%%%%%%%%%%%%%%%%%%%%%%

\end{document}